\documentclass[preprint,longbibliography,aps,preprintnumbers,amsmath,amssymb]{revtex4}
\usepackage{graphicx}
\usepackage{float}
\usepackage{bm}
\usepackage{amsmath}
\usepackage{epstopdf}
\usepackage{multirow}
\usepackage{footmisc}
\usepackage{inputenc}
\usepackage{lineno}
\usepackage[colorlinks=true, linkcolor=blue, citecolor=blue,urlcolor=blue]{hyperref}

\makeatletter

\@ifundefined{textcolor}{}
{%
 \definecolor{BLACK}{gray}{0}
 \definecolor{WHITE}{gray}{1}
 \definecolor{RED}{rgb}{1,0,0}
 \definecolor{GREEN}{rgb}{0,1,0}
 \definecolor{BLUE}{rgb}{0,0,1}
 \definecolor{CYAN}{cmyk}{1,0,0,0}
 \definecolor{MAGENTA}{cmyk}{0,1,0,0}
 \definecolor{YELLOW}{cmyk}{0,0,1,0}
}

\makeatother

\begin{document}

\title{Combined sub-sampling and analytical integration for efficient large-scale $GW$ calculations for 2D systems}

\author{Weiyi Xia$^{1}$}

\author{Weiwei Gao$^{2}$}

\author{Gabriel Lopez-Candales$^{1}$}

\author{Yabei Wu$^{3,4}$}

\author{Wei Ren$^{5}$}

\author{Wenqing Zhang$^{3,4}$}

\author{Peihong Zhang$^{1}$}
\email{pzhang3@buffalo.edu}

\affiliation{$^{1}$Department of Physics, University at Buffalo, State University of New York, Buffalo, New York 14260, USA}

\affiliation{$^{2}$ Center for Computational Materials, Oden Institute for Computational Engineering and Sciences, The University of Texas at Austin, Austin, TX 78712, USA}
 
\affiliation{$^{3}$ Department of Physics and Shenzhen Institute for Quantum Science \& engineering, Southern University of Science and Technology, Shenzhen, Guangdong 518055, China}

\affiliation{$^{4}$ Guangdong Provincial Key Lab for Computational Science and Materials Design, and Shenzhen Municipal Key Lab for Advanced Quantum Materials and Devices, Southern University of Science and Technology, Shenzhen, Guangdong 518055, China}

\affiliation{$^{5}$International Centre for Quantum and Molecular Structures and Department of Physics, Shanghai University, Shanghai 200444, China}

\date{\today}

\begin{abstract}
Accurate and efficient predictions of the quasiparticle properties of complex materials remain a major challenge 
due to the convergence issue and the unfavorable scaling of the computational cost with 
respect to the system size. Quasiparticle $GW$ calculations for two dimensional (2D) 
materials are especially difficult. The unusual analytical behaviors of the dielectric screening and the electron 
self-energy of 2D materials make the conventional Brillouin zone (BZ) integration approach rather inefficient 
and require an extremely dense $k$-grid to properly converge the calculated quasiparticle energies.
In this work, we present a combined non-uniform sub-sampling 
and analytical integration method that can drastically improve the efficiency of the BZ 
integration in 2D $GW$ calculations. Our work is distinguished from previous work in that, 
instead of focusing on the intricate dielectric matrix or the screened Coulomb interaction matrix, 
we exploit the analytical behavior of various terms of the convolved self-energy $\Sigma(\mathbf{q})$
in the small $\mathbf{q}$ limit. This method, when combined with another accelerated $GW$ method 
that we developed recently, can drastically speed-up (by over three orders of magnitude) $GW$ calculations for 2D materials. 
Our method allows fully converged $GW$ calculations for complex 2D systems at a fraction of computational cost,
facilitating future high throughput screening of the quasiparticle properties of 2D semiconductors for
various applications. To demonstrate the capability and performance of our new method, we have carried out fully converged 
$GW$ calculations for monolayer C$_2$N, a recently discovered 2D material with a large unit cell, 
and investigate its quasiparticle band structure in detail.
\end{abstract}
\maketitle

\section{Introduction}

Two dimensional (2D) materials are at the center of materials research in recent years. The intense research activities 
have resulted in the discovery of an impressive and growing list of 2D materials that were once considered rare and unstable. 
Among them, 2D semiconductors have received particular attention for their potential use in future electronics and energy related 
applications. With the increasing role that theory plays in the design and prediction of 2D semiconductors, the importance of accurate 
understanding of their electronic structures cannot be overstated. 
Although the $GW$ approximation \cite{GW1,GW2,GW3} has been recognized as one of the 
most accurate theories for predicting the quasiparticle properties of a wide range of materials, straightforward applications of the 
$GW$ method to 2D materials have been met with multiple computational challenges that make fully converged $GW$ 
calculations (even at the $G^0W^0$ level) rather difficult.
These challenges are so grave that, if not properly addressed, they may lead to false theoretical 
predictions and confusions. 
% For example, the reported values of the (direct) quasipartical band gap of monolayer MoS$_2$ 
% range from 2.19 to 2.97 eV [PRB 88, 245309 (2013)]
% or preventing the widespread adoption of the GW method for 2D materials predictions. 

One of the difficulties of 2D $GW$ calculations comes from the Brillouin zone (BZ) integration of the $GW$ self-energy, 
which is often carried out using discrete summation on a uniform $k$-grid ($N_1\times N_2\times1$ for 2D systems): 
\begin{equation}
\Sigma_{n\mathbf{k}}(\omega)=\frac{1}{\Omega}\int_{\Omega}\Sigma_{n\mathbf{k}}(\mathbf{q},\omega)d\mathbf{q}\approx
\sum_{\mathbf{q}}f_{\mathbf{q}}\Sigma_{n\mathbf{k}}(\mathbf{q},\omega),
\end{equation}
where $\Sigma_{n\mathbf{k}}(\mathbf{q},\omega)$ is the contribution to the GW self-energy for state 
$|n\mathbf{k}\rangle$ from point $\mathbf{q}$ in the BZ, $\Omega$ is the volume of
the BZ, and $f_{\mathbf{q}}$ is the appropriate weight.
This summation typically converges rather quickly with 
respect to the BZ sampling density for bulk (3D) semiconductors. 
For example, for silicon (diamond structure with a 2-atom unit cell), a $6\times6\times6$ $k$-grid is 
sufficient to converge the calculated $GW$ band gap to within 0.01 eV. For 2D materials, however, the convergence 
is extremely slow. It has been shown that one needs a $24\times24\times1$ $k$-grid to properly converge the $GW$ band gap of 
monolayer MoS$_2$ \cite{MoS21,MoS22,MoS23,2D3}.
Although this slow convergence issue is now well understood, it was somewhat unexpected at first.
% as evidenced by some earlier works that used very coarse $k$-grids. 
Since the computational cost of $GW$ calculations 
scales as $O(N_k^2)$, where $N_k$ is the number of the BZ integration points, 
the slow BZ integration convergence issue in 2D $GW$ calculations has significantly hindered
practical applications of the $GW$ method for accurate 2D materials predictions.

Compounding matters further is the need to include a large vacuum layer in the modeling of 2D systems
using the periodic supercell approach (to minimize the spurious interlayer interactions), resulting in a large cell volume 
even for relatively simple 2D materials with only a few atoms in the unit cell.
This is particularly true for theories (such as the $GW$ method) that involve the calculations of nonlocal 
interactions or response functions. The calculated quasiparticle energies converge extremely slowly with respect to
the vacuum layer thickness $d$ if unmodified long-range Coulomb interaction is used \cite{slab,Ismail-Beigi,coulomb2}.
Although the use of truncated Coulomb interaction \cite{Ismail-Beigi,coulomb2} 
greatly expedites the convergence with respect to $d$, the calculated results still depend on
the layer separation (albeit on a much weaker degree), and one still need to include
a sizable vacuum layer of about 20 \AA\ or greater for most 2D materials.

The large cell volume translates into the need to include a large number of electronic states in $GW$ calculations.
For example, it has been shown \cite{MoS21,MoS22,Gao2016,Wu_2018} that one may need to include up to 10,000 
conduction bands in the conventional $GW$ calculations even for simple 2D materials with a small unit cell of a few atoms. 
Note that in order to reach a similar level of convergence, this number scales linearly with the 
system size (i.e., number of atoms in the unit cell), making fully converged $GW$ calculations 
for more complex 2D systems extremely difficult using the conventional band-summation approach. 

Recently, we developed an accelerated $GW$ approach that can drastically speed up $GW$ calculations for large systems \cite{Gao2016}.
In this method, the computationally demanding band-summation in conventional $GW$ calculations is replaced by an 
energy-integration method, resulting in a speedup factor of up to two orders of magnitude for large and/or complex 
systems, including 2D materials. 
The slow BZ integration convergence issue, however, still poses a formidable challenge
for 2D $GW$ calculations. Considering the importance of accurate predictions of the quasiparticle properties of 2D materials, 
it is not surprising that there have been several proposed schemes that aim at addressing the slow BZ
integration convergence issue, noticeably the work of Rasmussen {\it et al.} \cite{2D2}
and that of da Jornada {\it et al.} \cite{2D3}.
Motivated by these works, we present here an efficient and accurate yet simple-to-implement method 
that can significantly reduce the required BZ sampling density for well converged 2D $GW$ calculations. 
We have tested our method for a range of 2D semiconductors \cite{Wu_2018,MXene}, 
and, for most cases, the calculated $GW$ quasiparticle energies
converge to within 50 meV or less using a very coarse  $6\times6\times1$  $k$-grid.
Combining these two new approaches, we are able to carry out fully converged 2D $GW$ calculations with an
overall speed-up factor of over three orders of magnitude compared with the conventional approach. 

% The rest of the paper is organized as follows: after a brief discussion of 
% the computational details in Sec. \ref{section:2}, we investigate in Sec. \ref{section:3}
% the analytical behavior of the contribution to the $GW$ self-energy of
% the valence (occupied) and conduction (unoccupied) states from near the BZ center using monolayer MoS$_2$ 
% as an example. The combined sub-sampling and analytical integration approach is proposed in 
% Sec. \ref{section:4}. Section \ref{section:5} demonstrates the performance of the new method using
% monolayer MoS$_2$ and C$_2$N as examples. Finally, we provide some concluding remarks in Sec. \ref{section:6}.

\section{Results}

\subsection{Analytical behavior of the GW self-energy of 2D systems}

\begin{figure*}
\includegraphics[width=0.95\textwidth]{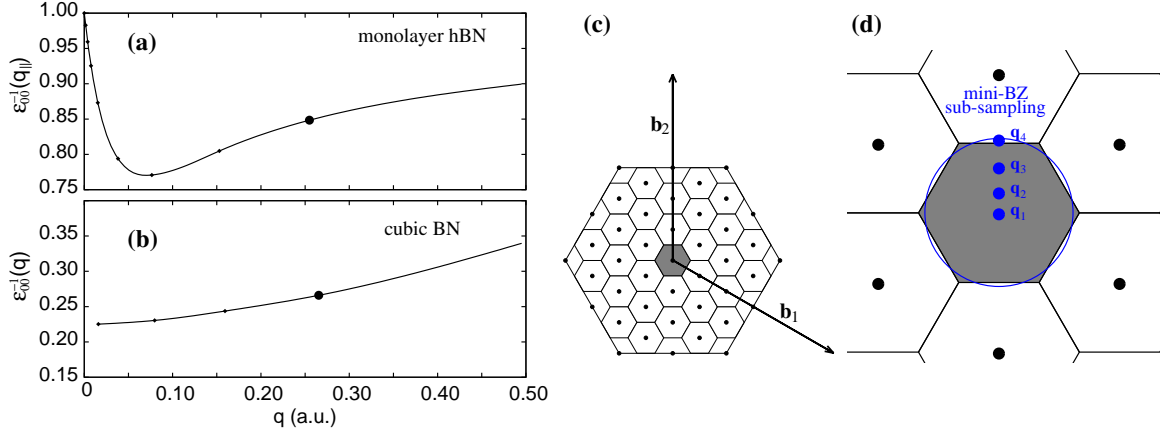}
\caption{{\bf Inverse dielectric functions of 2D and 3D boron nitride and sampling of a 2D BZ.}
The $q$-dependent head element of the inverse dielectric matrix of monolayer hexagonal boron nitride is shown in (a) and that
of cubic boron nitride in (b). 
The large black dots in (a) and (b) show the smallest $q$ (other than $q=0$) included in a $6\times6\times1$ (2D) or $6\times6\times6$ (3D) grid.
(c) A 2D hexagonal BZ with a $6\times6\times1$ uniform $k$-grid shown
with black dots. The gray-shaded area shows the mini-BZ enclosing the $\Gamma$ point.
(d) The mini-BZ with sub-sampling points indicated by blue dots.}
\label{fig:1} 
\end{figure*}

The slow BZ integration convergence issue in 2D $GW$ calculations is a manifestation of the asymptotic behavior
of $\Sigma_{n\mathbf{k}}(\mathbf{q},\omega)$ (defined in Eq. 1) in the long wavelength (small $q$) limit, which
is related to the analytical properties of the dielectric function  
$\epsilon^{-1}_{\mathbf{G}\mathbf{G'}}(\mathbf{q},\omega)$, or equivalently, that of
the screened Coulomb interaction $W_{\mathbf{G}\mathbf{G'}}(\mathbf{q},\omega)$.
These quantities vary rapidly as the wave vector $\mathbf{q}$ approaches zero, making
a simple discrete summation using the uniform sampling scheme very difficult to converge. 
If we write the BZ summation of the $GW$ self-energy into two parts,
\begin{equation}
\Sigma_{n\mathbf{k}}(\omega)=f_{0}\Sigma_{n\mathbf{k}}(\mathbf{q}=0,\omega)
+\sum_{\mathbf{q}\neq 0}f_{\mathbf{q}}\Sigma_{n\mathbf{k}}(\mathbf{q},\omega),
\end{equation}
it becomes clear that most convergence error comes from the $\mathbf{q}=0$ term, or, more precisely,
the contribution from the mini-BZ centered around the $\Gamma$ point.
{In fact, even in conventional $GW$ calculations using a uniform
$k$-grid, the contribution from the $\mathbf{q}=0$ term has to be 
treated carefully due to the divergence of the Coulomb interaction.
This is typically done by exploiting the analytical behavior of the dielectric matrix and
the (truncated) Coulomb interaction at the small $\mathbf{q}$ limit and carrying out
a mini-BZ averaging of the screened Coulomb matrix, as has been 
implemented in the BERKELEYGW package and has been discussed in great details in previous works \cite{GW2,Ismail-Beigi,GW4}.}

Figure \ref{fig:1} (a) and (b) compare the $q$-dependent head element $\epsilon_{00}^{-1}(q_\parallel)$ 
(here $q_\parallel$ denotes the wave vector parallel to the atomic plane of the 2D system) of 
the inverse dielectric matrix of monolayer hexagonal boron-nitride (hBN) and that of bulk cubic boron-nitride (cBN). 
The large black dots in the figure correspond a sampling point in a $6\times6\times1$ $k$-grid for the monolayer hBN and that
in a $6\times6\times6$ $k$-grid for bulk cBN. 
Whereas  $\epsilon_{00}^{-1}(q)$ of bulk cBN varies smoothly as $q$ approaches 0, due to the diminishing
2D dielectric screening in the long wave length limit, there is a sharp upturn of this quantity at small $q$ 
for the monolayer hBN system. Accurate capturing of such rapid variation
would require an extremely dense $k$-grid if uniform sampling schemes were used.
{Note that, strickly speaking, the dielectric function for a 2D system calculated 
using periodic boundary conditions is not a truly 2D dielectrc function but that of the 3D model system.
It has been shown \cite{Ismail-Beigi}, however, that if a truncated Coulomb potential is used, the calculated GW self-energy
converges quickly with increasing interlayer separation.}

Therefore, it is compelling to exploit the analytical behavior of the dielectric function and that of 
the screened Coulomb interaction to achieve converged $GW$ results without the need to use
a very high density BZ sampling grid.
Rasmussen {\it et al.} \cite{2D2} proposed a well-motivated analytical model for
the screened Coulomb interaction in the long wavelength limit for 2D systems and carried out
the integration of the self-energy in the mini-BZ centered around $\Gamma$.
Figure \ref{fig:1} (c) shows the BZ of a 2D hexagonal system. A $6\times6\times1$ uniform sampling 
grid is shown with black dots in the figure; the shaded area is the mini-BZ centered around the $\Gamma$ point.
Using this method, Rasmussen {\it et al.} \cite{2D2} showed that the 
calculated quasiparticle band gap of monolayer MoS$_2$ converges to about 0.1 eV
using  a $12\times12\times1$ \textit{k}-grid. Although this is a significant achievement, 
a $12\times12\times1$ $k$-grid is still fairly dense, and it is desirable to further reduce the required BZ sampling density.
Note that the computation cost of $GW$ calculations scales as $O(N_k^2)$, where $N_k$ is the number of the BZ integration points, a small reduction in the $k$-grid density will result in significant saving of the computation time.
For example, by reducing the 2D $k$-grid density from $12\times 12\times 1$ to $6\times 6\times1$, 
the computational cost would be reduced by a factor of 16.

Finding a compact and reliable analytic model for the the response function for
a wide range of 2D materials, even in the small $q$ limit, is difficult. 
Instead of exploiting the analytical behavior of the dielectric function, recently, da Jornada {\it et al.} 
\cite{2D3} proposed a non-uniform sub-sampling scheme to improve the quality of discreteel at
BZ integration. The screened Coulomb interaction matrix in the mini-BZ
is approximated by a weighted summation of a few sub-sampling points
in the mini-BZ as shown schematically in Fig. \ref{fig:1} (d). Using the method, it was shown that the quasiparticle band gap 
of bilayer MoSe$_2$ converges to within 50 meV using a coarse 2D $k$-grid of $6\times 6\times1$ 
with 10 sub-sampling points in the mini-BZ.  
In both of these methods, a better convergence is achieved with more accurate evaluation of
the rapid variation of the screened Coulomb interaction matrix $W_{\mathbf{G}\mathbf{G'}}(\mathbf{q})$ within the mini-BZ. 
However, instead of working on the screened Coulomb interaction matrix,  we believe that it is more efficient 
to exploit the analytical behavior of $q$-dependent self-energy contribution $\Sigma_{n\mathbf{k}}(\mathbf{q})$ directly.

\begin{figure*}
\includegraphics[width=0.95\textwidth]{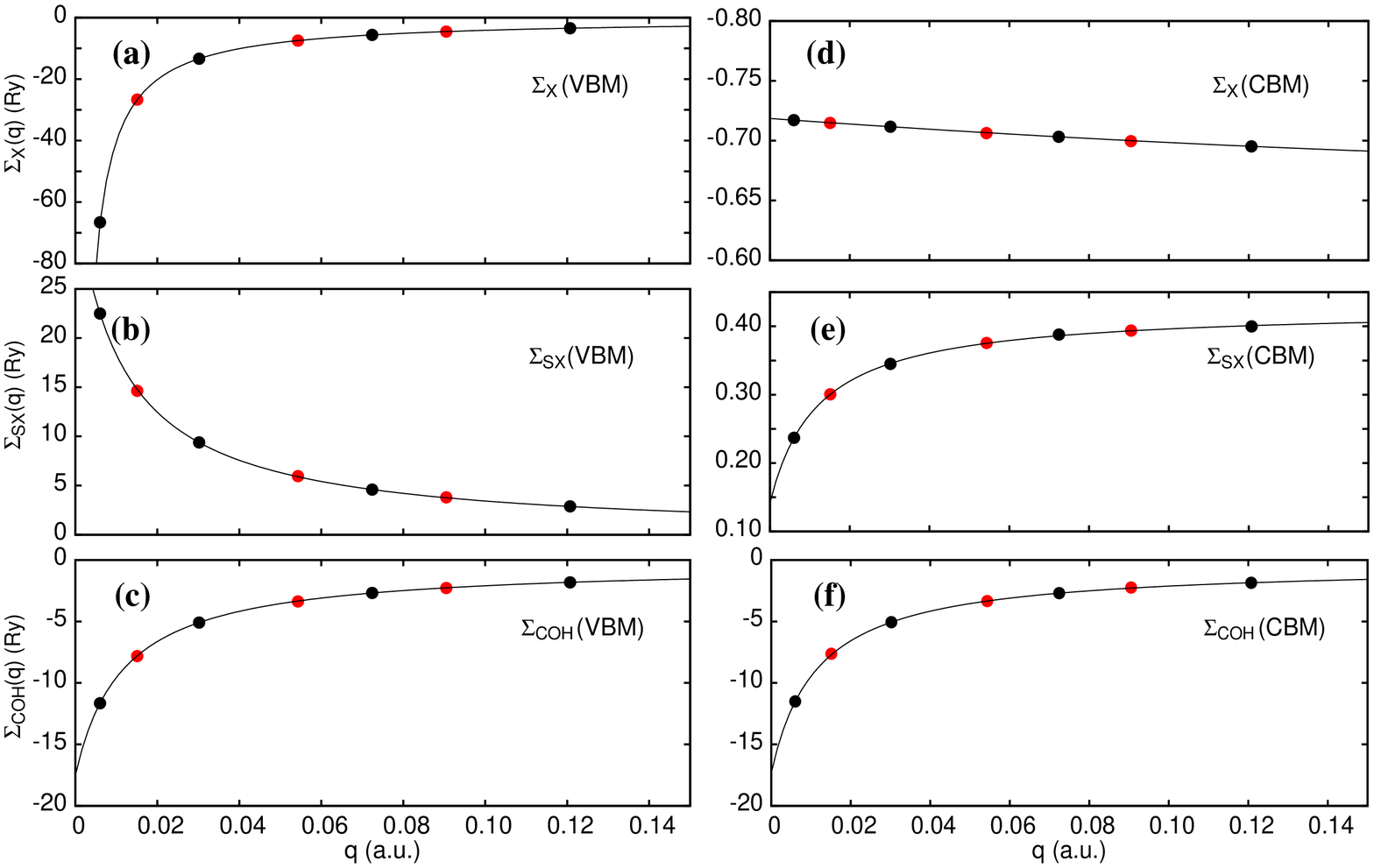}
\caption{{\bf The $q$-dependent contribution to the self-energy of monolayer MoS$_2$.}
The $q$-dependent contributions to the self-energy of the VBM are shown in the left panels, 
and that of the CBM are shown in the right panels.
The self-energy is decomposed into three terms of different physical origins as discussed in the text.
The large black dots indicate the calculated values at the sub-sampling $q$ points, and
 the curves show the fitting functions. The red dots show results calculated at a few additional $q$ points, which
 agree extremely well with the fitting functions.}
 \label{fig:2} 
\end{figure*}

The electron self-energy can be conveniently separated into two parts, a screened exchange ($\Sigma^{\mathrm{SEX}}$) and a
Coulomb hole ($\Sigma^{\mathrm{COH}}$) part \cite{GW2}; the screened exchange part can be further
separated into a bare exchange ($\Sigma^{\mathrm{X}}$) and a correction term  ($\Sigma^{\mathrm{SX}}$) arising from
the screening potential:
\begin{equation}
\Sigma=\Sigma^{\mathrm{SEX}}+\Sigma^{\mathrm{COH}}=(\Sigma^{\mathrm{X}}+\Sigma^{\mathrm{SX}})+\Sigma^{\mathrm{COH}}.
\end{equation}
Figure \ref{fig:2} shows these self-energy terms (solid dots) for the valence band maximum (VBM, left panels) and 
conduction band minimum (CBM, right panels) states of monolayer MoS$_2$ as a function of wave vector $q$.
Note that it is the integration of these contributions over the BZ that gives the self-energy correction for
the electronic state (e.g., VBM or CBM) of interest. 
Interestingly, all these quantities show well-behaved asymptotic properties.
Therefore, it is very important to analyze and exploit the analytical behavior of these quantities in the small $q$ limit.

We first examine the screened exchange energy for state $|n\mathbf{k}\rangle$ :
\begin{equation}
\Sigma_{n\mathbf{k}}^{\mathrm{SEX}}(\omega)=\sum_{v\mathbf{q},\mathbf{GG'}}M_{nv}^{*}(\mathbf{k},\mathbf{q},\mathbf{G})
M_{vn}(\mathbf{q},\mathbf{k},\mathbf{G'}) W_{\mathbf{GG'}}(\mathbf{q},\omega),
\end{equation}
where $M_{vn}(\mathbf{q},\mathbf{k},\mathbf{G})=\langle v,\mathbf{k+q}|
e^{i(\mathbf{q}+\mathbf{G})\cdot\mathbf{r}}|n,\mathbf{k}\rangle$ 
are plane-wave matrix elements between the two states
$|v,\mathbf{k+q}\rangle$ and $|n,\mathbf{k}\rangle$.
It is convenient to write the screened Coulomb
potential as the summation of the bare Coulomb potential $v_\mathrm{b}$ and the
screening potential $v_{\mathrm{scr}}$, i.e., $W=v_b+v_{\mathrm{scr}}$.
Correspondingly, the screened exchange energy can be separated into the bare exchange
and a correction coming from the screening potential, i.e., 
$\Sigma^{\mathrm{SEX}}=(\Sigma^{\mathrm{X}}+\Sigma^{\mathrm{SX}})$, as mentioned earlier.
 
The analytical behavior of the bare exchange energy $\Sigma^{\mathrm{X}}(\mathrm{q})$
can then be understood by examining the truncated 2D Coulomb potential \cite{Ismail-Beigi} in
the momentum space:
\begin{equation}
v_{\mathbf{G}}^{2D}(\mathbf{q}_{\parallel})=\frac{4\pi}{|\mathbf{q}_{\parallel}+\mathbf{G}|^{2}}
\left[1-e^{-|\mathbf{q}_{\parallel}+\mathbf{G}_{\parallel}|L_z/2}\cos(\mathbf{G}_{z}L_z/2)\right],
\end{equation}
where $\mathbf{q}_{\parallel}$ is the wave vector within the 2D BZ,  $L_z$ is periodicity along the $z$ direction, 
$\mathbf{G}_\parallel$ ( $\mathbf{G}_z$) denotes the $\mathbf{G}$ vectors that are parallel (perpendicular)
to the 2D atomic layer. The truncated Coulomb potential then approaches $2\pi L/|\mathbf{q}_{\parallel}|$ in the small $q$ limit. For simplicity, we will drop the parallel sign ($\parallel$) for wave vectors $q$ within the 2D BZ if there are no
confusions. Therefore, it is straightforward to speculate that the leading term of the bare exchange energy
for the valence (occupied) states has the same asymptotic expression as the bare Coulomb potential. Extending the
expression to finite $q$, for 2D isotropic systems, we have
\begin{equation}
\Sigma^{\mathrm{X}}_{v\mathbf{k}}(q)\approx\frac{A}{q}+B+Cq.
\end{equation}

The solid curve of the top-left panel of Fig. \ref{fig:2} shows a perfect 3-parameter fitting of 
$\Sigma^{\mathrm{X}}_{\mathrm{VBM}}(q)$ of monolayer MoS$_2$ calculated on four $q$ points 
indicated with large black dots. With this fitted expression, 
the integration of $\Sigma^{\mathrm{X}}_{n\mathbf{k}}(q)$ within the mini-BZ can be carried out
analytically. Due to the absence of the self-exchange,
the bare exchange for conduction (unoccupied) states is much smaller than that for occupied states, and is basically featureless,
as shown in the top-right panel of Fig. \ref{fig:2}, which can be well fitted with a 2- or 3-parameter function, i.e.,
\begin{equation}
\Sigma^{\mathrm{X}}_{c\mathbf{k}}(q)\approx A+Bq+Cq^2,
\end{equation}
using values calculated on 4 $q$ points as shown with solid curve in the top-right panel of Fig. \ref{fig:2}.

The correction to the exchange energy arising from the dielectric screening of
the Coulomb potential can also be analyzed. The 2D dielectric function takes the form 
\cite{2ddielectric,2D2}
\begin{equation}
\epsilon(q)\approx1+2\pi\alpha_{\mathrm{2D}}q
\end{equation}
in the long wavelength limit, where $\alpha_{\mathrm{2D}}$ is the 2D polarizability.
Therefore, the screening potential $v_{\mathrm{scr}}$ takes the form 
$v_{\mathrm{scr}}\approx4\pi^2\alpha_{2D}/(1+2\pi\alpha_{\mathrm{2D}}q)$.
Considering the dynamical screening effects, we propose the following analytical form for 
$\Sigma^{\mathrm{SX}}_{n\mathbf{k}}(q,\omega)$ for both valence and conduction states:
\begin{equation}
\Sigma^{\mathrm{SX}}_{n\mathbf{k}}(q,\omega)\approx\frac{A(\omega)}{1+B(\omega)q}+C(\omega).
\end{equation}
The middle panels of Fig. \ref{fig:2} show the 3-parameter fittings for 
$\Sigma^{\mathrm{SX}}(q)$ for the VBM and CBM states of monolayer MoS$_2$
calculated at their respective DFT energies.
Finally, we find that the Coulomb hole self-energy can also be well fitted with the same analytical form, i.e.,
\begin{equation}
\Sigma^{\mathrm{COH}}_{n\mathbf{k}}(q,\omega)\approx\frac{A(\omega)}{1+B(\omega)q}+C(\omega),
\end{equation}
for both the valence and conduction states as shown in the bottom panels
of Fig. \ref{fig:2}. 

We have implemented a nonlinear fitting algorithm (the Levenberg-Marquardt algorithm)
in our code. We monitor the fitting quality, i.e., the residual error, so one can easily spot 
possible issues with the fitting procedure. For all systems we have studied, the fitting procedure converges quickly
with a reasonable initial guess (e.g., by setting all inititial parameters to 1.0).
In order to demonstrate the reliability and quality of the proposed fitting functions and that of the implemented
fitting algorithm, we have calculated the self-energy at a few additional $\mathbf{q}$ points and 
have added these data points (red dots) to Fig. \ref{fig:2}.
These additional data points agree well with the functions fitted using the original data (black dots).

\subsection{Combined sub-sampling and analytical integration approach}

Putting these results together, we propose an approach
that have the advantages of both of the previously proposed schemes 
\cite{2D2,2D3}, a combined
sub-sampling and analytical integration of the self-energy within the mini-BZ,
to tackle the convergence issue of the BZ integration in 2D $GW$ calculations.
The BZ is sampled with a coarse uniform $k$-grid as usual; a $6\times6\times1$ 
grid is sufficient for most 2D systems with small unit cells. For complex
2D materials with large unit cells, an even coarser $k$-grid may be used as we will
discuss later. We then carry out a few additional sampling points inside the mini-BZ.
The three $q$-dependent $GW$ self-energy terms, namely,
$\Sigma^{\mathrm{X}}(\mathbf{q})$,
$\Sigma^{\mathrm{SX}}(\mathbf{q},\omega)$, and
$\Sigma^{\mathrm{COH}}(\mathbf{q},\omega)$
are calculated on these additional sampling points and the results are fitted using
the analytical functions discussed in the previous section.
The BZ integration of the $GW$ self-energy is separated into two parts,
a conventional weighted summation over all $k$-points except the $\Gamma$ point, and
an integration of the fitted analytical functions over the mini-BZ:
\begin{equation}
\Sigma_{n\mathbf{k}}(\omega)=\frac{f_0}{\Omega_0}\int_{\Omega_0}\Sigma_{n\mathbf{k}}
(\mathbf{q},\omega)d\mathbf{q}+\sum_{\mathbf{q}\neq 0}f_{\mathbf{q}}\Sigma_{n\mathbf{k}}(\mathbf{q},\omega),
\end{equation}
where $\Omega_0$ is the area of the 2D mini-BZ as shown in Fig. \ref{fig:1} (d). 

The self-energy $\Sigma_{n\mathbf{k}}(\mathbf{q},\omega)$ is tyipcally calculated at two energy points,
$\omega=\epsilon_{n\mathbf{k}}^{\mathrm{DFT}}$, and $\omega=\epsilon_{n\mathbf{k}}^{\mathrm{DFT}}+\Delta\epsilon$.
A linear expansion  \cite{GW2} of the self-energy is then carried out to obtain the self-energy evaluated at the quasiparticle energy, i.e.,
 $\Sigma_{n\mathbf{k}}(\omega=E^{\mathbf{QP}}_{n\mathbf{k}})$.
Since the integration over the mini-BZ 
is carried out using the fitted analytical functions as opposed to the weighted summation approach, 
we need only a small number of sub-sampling points. 
In fact, for all isotropic 2D systems we have studied, 4 additional sampling points are sufficient 
to converge the calculated quasiparticle energy to within 0.01 eV for a given $N\times N\times1$
BZ sampling grid, as discussed in the next section. We mention that our method can be extended to treat anisotropic 2D materials.
In this case, the sub-sampling calculations within the mini-BZ have to be carried out
along the two reciprocal lattice directions $\mathbf{b}_1$ and $\mathbf{b}_2$, and some of the fitting
coefficients are vectors instead of scalars.  
We will report results for anisotropic 2D systems in a separate publication.

\subsection{Convergence behavior of the GW band gap of monolayer MoS$_2$}

\begin{figure}[h]
\includegraphics[width=0.48\textwidth]{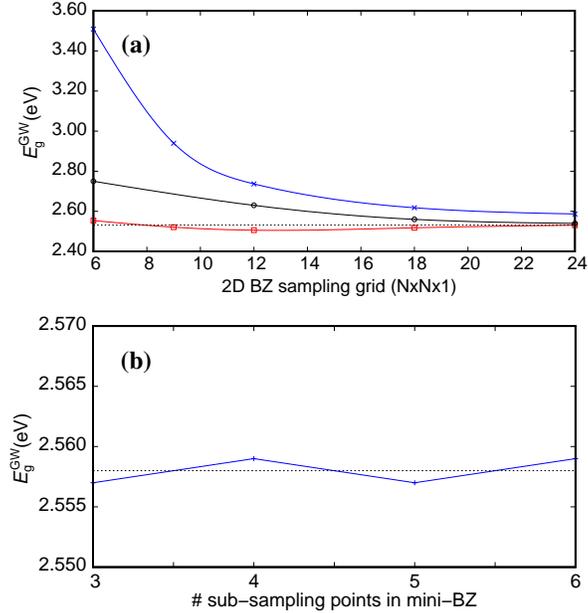}
\caption{
{\bf Convergence behavior of the quasiparticle band gap of MoS$_2$.}
(a) Calculated quasiparticle band gap of MoS$_2$ with respect to increasing BZ
sampling density using the conventional uniform sampling 
method (blue curve) and the current method (red curve).
A very coarse $6\times6\times1$ $k$-grid with four additional sub-sampling points in the mini-BZ is
sufficient to converge the calculated band gap to within 0.02 eV using our method. 
The dotted line is guide for the eye, showing the converged value. We also include the results 
of Rasmussen {\it et al.} \cite{2D2} (black solid curve) for comparision.
(b) Calculated band gap of MoS$_2$ using different number of sub-sampling $q$ points in the mini-BZ. 
The extremely small variation ($\pm$ 2 meV) likely comes from numerical errors,
suggesting that our calculations essentially converge with as few as 3 sub-sampling $q$ points.}
\label{fig:3} 
\end{figure}

We first demonstrate the performance of our method using monolayer MoS$_2$ as an example.
The quasiparticle properties of monolayer MoS$_2$ have been investigated by several groups \cite{MoS21,MoS22,MoS23,MoS24,MoS25, Gao2016}. 
Therefore, this system serves as a good model for testing our methods.
Figure \ref{fig:3} (a) shows the calculated minimum direct gap at the $K$ point of monolayer MoS$_2$ as a
function of the $k$-point sampling density. Using the uniform sampling
approach, the calculated $GW$ band gap converges to within 0.05 eV with a very dense $24\times24\times1$ $k$-grid.
Using our approach, the band gap converges to within about 0.02 eV with a $6\times6\times1$ grid.
Note that the spin-orbital coupling effects are not included in the results shown in the figure.
As mentioned earlier, the computational cost of the dielectric matrix scales as $O(N_k^2)$, where
$N_K$ is the number of the BZ sampling points. Reducing the $k$-grid density from 
 $24\times24\times1$ to $6\times6\times1$ would ideally result in a speed-up factor of 256.
We achieve a speed-up factor of about 200 in real calculations, including the overhead
associated with the calculation of the the four sub-sampling $q$ points in the mini-BZ.

We include in Fig. \ref{fig:3} (a) the results of Rasmussen {\it et al.} \cite{2D2} (black solid curve) for comparision. 
It should be mentioned that our result seems to agree with that of Rasmussen {\it et al.} \cite{2D2} calculated with a
24$\times$24$\times$1 $k$-grid. This is a coincidence rather than a confirmation considering various differences 
(e.g., pseudopotential, crystal structure, and several cutoff parameters) in the two calculations.

We have tested the calculated band gap with respect to the number of sub-sampling points in the mini-BZ, as shown in
Fig. \ref{fig:3} (b). The result essentially converges with 3 sub-sampling $q$ points.
The extremely small error ($< 5$ meV) likely comes from numerical errors instead of from the systematic convergence error.
We have also tested the sensitivity of the results on the choice of the sub-sampling $q$-points, and
we can confirm that the results are fairly insensitive. Different choices of 
the sub-sampling $q$-points within a given $N\times N\times 1$
$k$-grid give practical identical results; the difference is usually within a few meV.
This is expected since the calculated self-energy can be fitted extremely well with
the proposed functional forms as shown in Fig. \ref{fig:2}.

\begin{figure}[h]
\includegraphics[width=0.48\textwidth]{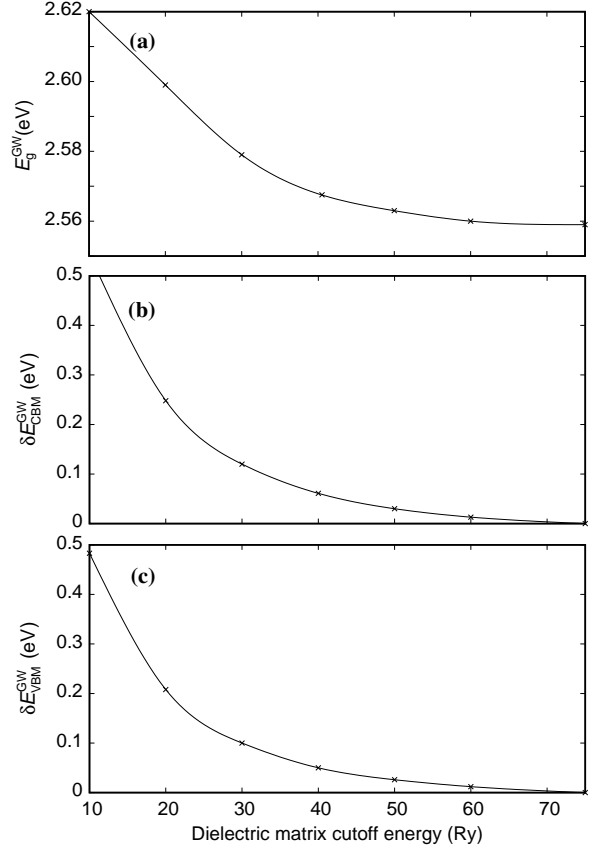}
\caption{ {\bf Convergence behavior of the quasiparticle band gap and energies of MoS$_2$.}
Calculated quasiparticle band gap (a), error in the quasiparticle energy of the 
CBM state (b), and that of the VBM state (c) of MoS$_2$ as a function of
the cutoff energy of the dielectric matrix.
}
\label{fig:4} 
\end{figure}

Although the main focus of this work is to address the slow convergence issue of the BZ integration
in GW calculations for 2D materials, we would like to discuss other convergence issues in GW calculations.
These issues may become another bottleneck for GW calculations for 2D materials.
Fig. \ref{fig:4} (a) shows the calculated band gap of MoS$_2$ (without including the spin-orbit coupling effects)
as a function of the cutoff energy of the dielectric matrix. The calculated band gap does not seem to show a significant
dependence on this cutoff parameter, decreasing from 2.62 eV to 2.56 eV when the cutoff energy is increased from 10 Ry to 75 Ry. 
A closer look at the convergence behavior of the quasiparticle energies of
the VBM and CBM states, however, reveals a rather different picture.
The CBM energy decreases by over 0.5 eV, whereas the VBM energy decreases by
slightly less than 0.5 eV, within the same parameter range, as shown in Fig. \ref{fig:4} (b) and (c).
Both the VBM and CBM of MoS$_2$ are primarily derived from the Mo $d$ states,
these states share similar wave function characteristics, thus similar convergence
behavior. Therefore, the errors largely cancel out, making the the calculated band gap appears to depend 
only weakly on this cuoff parameter. 
However, if the states of interest have significantly different wave function characteristics, 
highly converged calculations are necessary, and under-converged calculations may give
false predictions for important properties such as transition energies of band offsets.

\begin{figure}[h]
\includegraphics[width=0.48\textwidth]{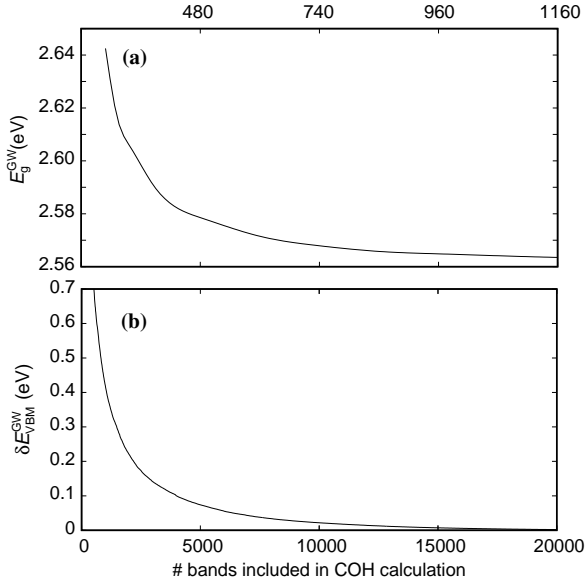}
\caption{{\bf Convergence behavior of the quasiparticle band gap and energies of MoS$_2$.}
Calculated $GW$ band gap (a) and and the error in quasiparticle energy of the VBM state (b) as
a function of the number of bands included in
the calculation. The lower horizontal axis shows the effective number of bands 
included in our $GW$ calculations whereas the
upper horizontal axis shows the actual number the integration
points using our method \cite{Gao2016}.
}
\label{fig:5} 
\end{figure}

Our calculations also benefit from the energy-integration method \cite{Gao2016,Gao2018} we developed 
to speed up the band summation in $GW$ calculations. As we have mentioned earlier,
conventional GW calculations for 2D materials require to include a large number of 
conduction bands, making highly converged calculations even for simple 2D materials containing a few atoms very difficult.
The total number of the empty states in our calculations for the monolayer MoS$_2$ is about 25,000, 
and one needs to include about 10,000 bands to properly converge the band gap
(to within 0.01 eV) as shown in Fig. \ref{fig:5} (a). 
Using our energy integration method, we only need about 740 integration (sampling) points
to achieve the same level of convergence. 
Similar to what we have discussed earlier, the change in the calculated band gap
with respect to the number of bands included in the GW calculations is
much smaller than the change in the VBM (or CBM) quasiparticle energy due to error cancellation,
as shown in Fig. \ref{fig:5} (b). However, there are situations in which highly converged
results for the quasiparticle energyies (not just the band gap) are required.
One of the advantages of our method is that we can afford to include (effectively) all empty states in our GW 
(both for the dielectric matrix and the self-energy) calculations without the need to concern about
the band summation convergence issue.

\subsection{Quasiparticle band structure of monolayer C$_2$N}

\begin{figure}[h]
\includegraphics[width=0.45\textwidth]{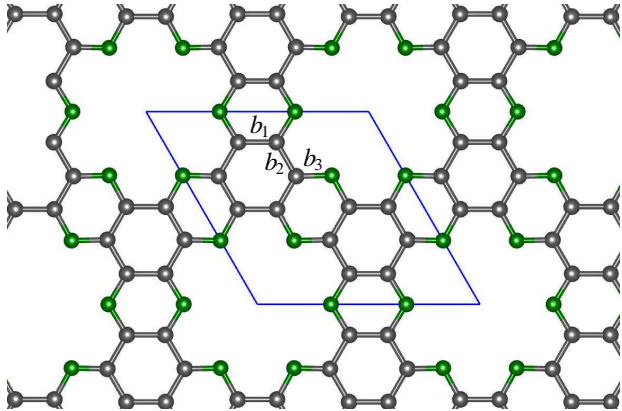}
\caption{{\bf Crystal structure of monolayer C$_2$N.}
 The grey and green balls represent C and N atoms, respectively.
The three unique bond lengths are $b_1=1.423$ \AA, $b_2=1.462$ \AA, and $b_3=1.331$ \AA.}
\label{fig:6} 
\end{figure}

In order to further demonstrate the capability and performance of our method, we now investigate 
the quasiparticle band structure of C$_2$N \cite{C2N1}, an interesting 2D carbon nitride that 
is distinguished from other 2D systems by its unique {\it holey} structure as shown in
Fig. \ref{fig:6}. 
The theoretically optimized lattice constant is 8.29 \AA; the three 
unique bond lengths are shown in the figure.
The structure has a large unit cell of 18 atoms, making fully converged $GW$
calculations a real challenge. In fact, C$_2$N has a 2D unit cell area
that is equivalent to that of a 24-atom graphene supercell.

\begin{figure}[h]
\includegraphics[width=0.49\textwidth]{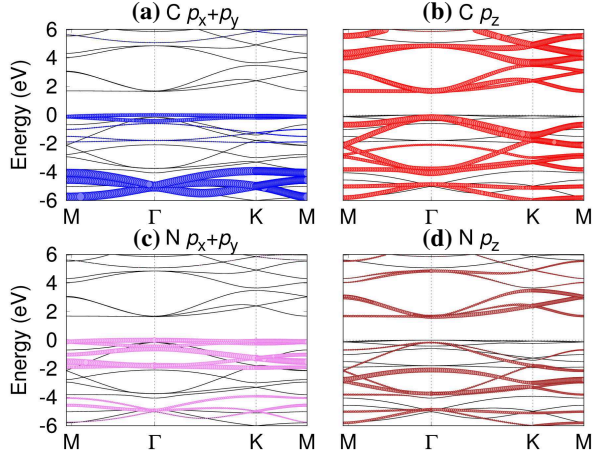}
\caption{
{\bf Projected DFT band structure of monolayer C$_2$N.}
The Bloch wave functions are projected onto different atomic orbitals to show
the distinct characters of the low energy valence and conduction states.}
\label{fig:7}
\end{figure}

The basic electronic structure of monolayer C$_2$N has been studied by several groups
\cite{C2N3,C2N4,C2N5,C2N6}.
One interesting feature of the band structure of monolayer C$_2$N is that
the top valence bands are nearly dispersion-less if local or semilocal energy functionals within DFT are used.
These flat valence bands are primarily derived from nitrogen and carbon $p_x$ and $p_y$ orbitals as
shown in Fig. \ref{fig:7} (a) and (c). The two valence bands immediately below the top
two (at the $\Gamma$ point), in contrast, are mostly derived from carbon $p_z$ orbitals with
small nitrogen $p_z$ components as shown in Fig. \ref{fig:7} (b) and (d).
Since the in-plane ($p_x$ and $p_y$) states may experience significantly different quasiparticle
self-energy corrections compared with the out-of-plane $p_z$ states, 
the ordering of these closely spaced valence bands may change
after including $GW$ self-energy corrections, which would have
important consequences on the calculated optical and transport properties of this material.
In the following, we first discuss the converged quasiparticle band structure of monolayer 
C$_2$N and discuss its important features compared with that calculated 
using the LDA. We then discuss several important convergence issues of the GW results.

\begin{figure}[h]
\includegraphics[width=0.48\textwidth]{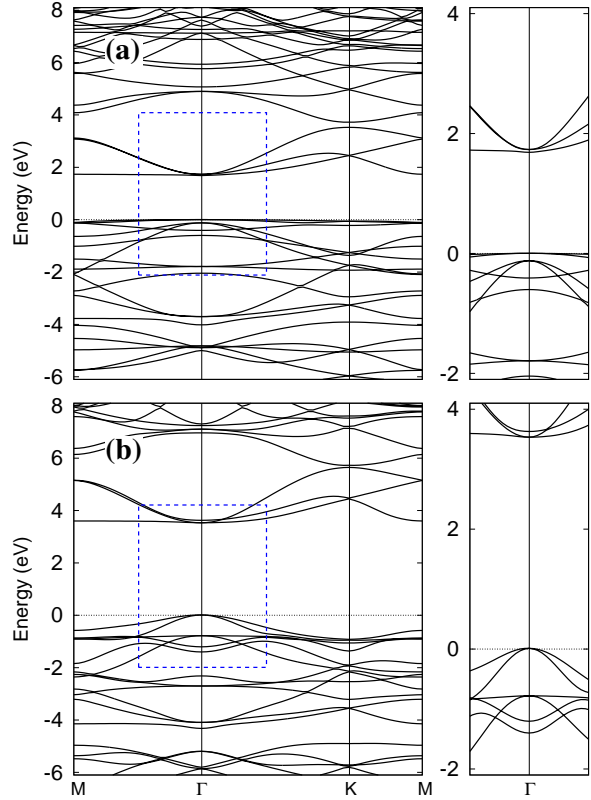}
\caption{{\bf Band structures of monolayer C$_2$N.}
The band structures are calculated with the LDA (a) and the $GW$ (b) methods. 
The areas indicated by blue rectangles are enlarged and shown in the right panels
to better illustrate changes in the band ordering after including the $GW$ corrections.}
\label{fig:8} 
\end{figure}

\begin{figure}[h]
\includegraphics[width=0.48\textwidth]{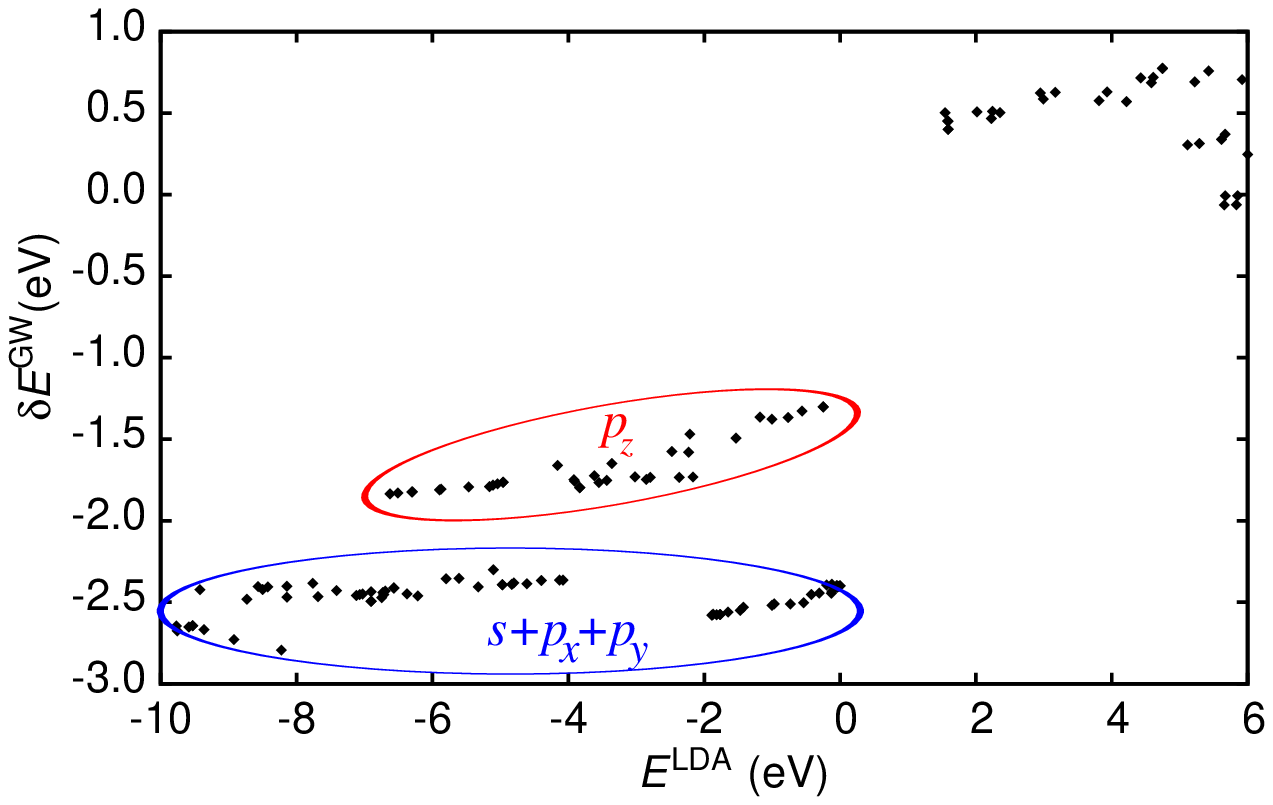}
\caption{$GW$ Quasiparticle corrections v.s. DFT energies plot showing distinct quasiparticle corrections for states with different atomic characters.}
\label{fig:9} 
\end{figure}

Figure \ref{fig:8} compares the DFT-LDA and the $GW$ band structures of monolayer C$_2$N.
As we have mentioned earlier, the LDA band structure shows two extremely flat top valence bands
which are derived from the in-plane carbon and nitrogen orbitals ($p_x$ and $p_y$)
as shown in Fig. \ref{fig:7}. The valence bands immediately below the two flat bands
are significantly more dispersive and are derived mostly from the
out-of-plane carbon $p_z$ orbitals.
The $GW$ band structure, on the other hand, shows rather dispersive top valence bands.
Upon a closer inspection, we find that this difference in the top valence band dispersion
comes from the contrasting $GW$ corrections to the out-of-plane ($p_z$) and in-plane states ($p_x$ and $p_y$)
as shown in Fig. \ref{fig:9}. The $p_x$ and $p_y$ derived valence states have significantly 
larger self-energy corrections compared with those of $p_z$ derived states.
As a result, the flat top-most valence states calculated within the LDA
drop below the $p_z$ derived states after including the $GW$ correction.
The $p_z$ derived states become the top-most valence states and are more dispersive.

To better illustrate the change in the band ordering, we show the
zoomed-in band structure around the $\Gamma$ point in the right panels of
Fig. \ref{fig:8}. We note that a similar valence band ordering change was observed 
earlier \cite{C2N5} with the use of HSE06 hybrid functional \cite{HSE1,HSE2}.
Interestingly, we find that the band ordering change also occur to the conduction bands
(although not as significant as that of valence bands) as shown in the right panels of Fig. \ref{fig:8}. 
These changes in the ordering of the band edge state will have profound impact
of the calculated optical and transport properties of this material, which deserve further
investigations.

\begin{figure}[ht]
\includegraphics[width=0.48\textwidth]{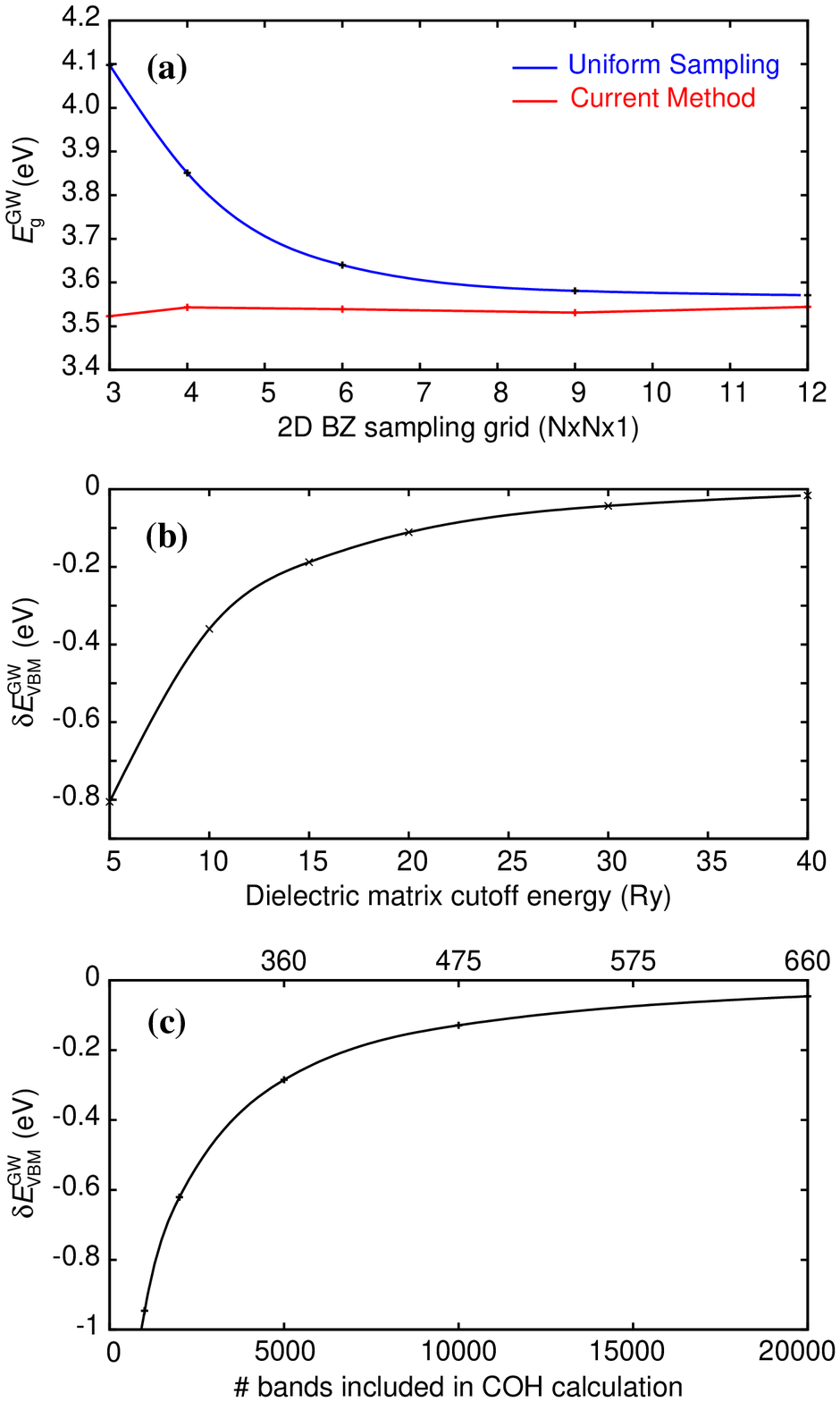}
\caption{{\bf Convergence behavior of the quasiparticle band gap and energy of monolayer C$_2$N.}
(a) The calculated GW band gap with respect to BZ sampling density using
the conventional uniform sampling method (blue curve) and the method (red curve) proposed in this work.
Error in the calculated quasiparticle energy of the VBM state as a function of
the cutoff energy of the dielectric matrix (b) and the number of conduction
bands included in calculation of the Coulomb hole (COH) self-energy (c)
The lower horizontal axis in (c) shows the number of bands 
to be included in conventional $GW$ calculations whereas the
upper horizontal axis shows the number of bands plus the integration
points used in our method \cite{Gao2016} to achieve the same
level of convergence.}
\label{fig:10} 
\end{figure}

We now discuss several important convergence issues of $GW$ calculations of this material.
Figure \ref{fig:10} (a) compares the calculated direct band gap as a function of of the BZ integration $k$-point 
density using the uniform sampling approach and the current method. 
Due to its relatively large unit cell (thus a small BZ), the calculated band
gap converges to within 0.02 eV using a very coarse $3\times3\times1$ $k$-grid,
or within 0.01 eV using a $4\times4\times1$ $k$-grid, with our BZ integration
method.  Quasiparticle $GW$ calculations of monolayer C$_2$N have been reported earlier\cite{C2N4}.
The authors used a very small cutoff energy (5 Ry) for the 
dielectric matrix and included only a few hundred bands in the calculations of
the dielectric matrix and the self-energy. The reported $GW$ band gap of monolayer C$_2$N 
was 3.75 eV \cite{C2N4}, to be compared with our result of 3.54 eV.

As we have discussed earlier, for many systems, the calculated $GW$ band gap may appear to converge while
the absolute quasiparticle energies for the valence and conduction bands are still not converged.
This is because the valence and conduction bands may have the similar convergence behavior, and
their difference (which defines the band gap) may appear to converge quickly.
In fact, a fairly high kinetic energy cutoff for the dielectric matrix and a large amount
of conduction bands are still needed in this case to achieve highly converged results for
the quasiparticle energies of this system.  

Figure. \ref{fig:10} (b) shows the convergence behavior of the calculated
quasiparticle energy for the VBM state as a function of the kinetic cutoff for the
dielectric matrix. If a 5 Ry dielectric matrix cutoff were used, the error in
the quasiparticle energy would be about 0.8 eV. A fairly high cutoff
energy of 30 Ry is needed to converge the quasiparticle energy to within 0.05 eV.
Figure \ref{fig:10} (c) shows the convergence behavior of the calculated
quasiparticle energy for the VBM state as a function of the number of conduction bands
included in the $GW$ calculations. Over 20,000 bands are needed to converge
the calculated quasiparticle energy to within 0.05 eV due to the large cell
size of this system. The error in the calculated quasiparticle energy is about 0.95 eV 
if 1,000 bands are included in the calculation.
Using the energy integration method that we developed \cite{Gao2016,Gao2018},
we are able to drastically reduce the computation cost associated with the
band summation in $GW$ calculations. The values shown on the lower horizontal axis are the
number of bands and integration grid points used in our calculations,
which shows a speed-up factor of about 30 (20000/660). Combining this method with
the non-uniform BZ integration method discussed in this work, we have achieved
a speed-up factor of well over three orders of magnitude.

\section{Methods}

We use the crystal structures optimized using the Perdew-Burke-Ernzerhof (PBE) functional \cite{PBE}
for subsequent electronic structure calculations. The optimized lattice constant for MoS$_2$ is  3.18 \AA,
and the layer thickness (i.e., the S-S interlayer distance) is 3.16 \AA. These values
are in reasonable agreement with published theoretical results. The detail of the
crystal structure of C$_2$N will be discussed later. 
The monolayer systems are modeled with periodic cells with an interlayer
separation of 25 \AA.
The mean-field electronic structure calculations are carried out using 
the pseudopotential plane-wave-based density functional theory (DFT) method within the
local density approximation (LDA) as implemented in
a local version of the PARATEC package \cite{paratec1,paratec2,paratec3}.
The Perdew-Zunger \cite{PZ-CA} parametrization of the Ceperley-Alder result \cite{CA} for the electron
correlation energy is used.
% PARATEC, http://www.nersc.gov/users/software/applications/materials-science/paratec/.
We use the Troullier-Martins norm-conserving pseudopotential \cite{Troullier}.
Semicore 4$s$ and 4$p$ of Mo are included in the calculation. 
The plans wave cutoff for the DFT and GW calculations for MoS$_2$ is set at
125 Ry; for C$_2$N, it is 70 Ry.

The $GW$ quasiparticle calculations are carried out within the $G^0W^0$  (i.e., one-shot $GW$) approach
\cite{GW2} using a local version of the BERKELEYGW package 
\cite{GW4} in which the method described
in this work and a recently developed energy-integration method \cite{Gao2016,Gao2018} 
are implemented. The summation over the conduction bands in $GW$ calculations is carried out
using the energy-integration approach \cite{Gao2016,Gao2018}. Using this
method, we can effectively include {\it all} conduction bands in the calculations at a fraction of the
computational cost compared with the conventional band-by-band summation. 
The kinetic energy cutoff for the dielectric matrix is set at 75 Ry for MoS$_2$ and 40 Ry for C$_2$N.
These cutoffs are sufficient to converge the calculated quasiparticle band gap to within 0.02 eV.
We use the Hybertsen-Louie generalized plasmon-pole model (HL-GPP)\cite{GW2}
to extend the static dielectric function to finite frequencies.

\section{Discussion}

Accurate and efficient $GW$ calculations for 2D materials are met with a multitude of computational challenges. 
The computational cost of fully converged $GW$ calculations for 2D materials, even for simple materials with
small unit cells of a few atoms, can be very expensive, making reliable $GW$ calculations for large and/or 
complex 2D systems a daunting task. The formidable computational demand has significantly held back the 
widespread adoption of this otherwise highly successful method for 2D materials predictions. 

By carefully investigating the analytical 
behavior of the $GW$ self-energy, we proposed a combined sub-sampling and analytical integration method that can
greatly improve the efficiency of 2D $GW$ calculations, enabling fast and accurate quasiparticle calculations for complex 2D systems. For most simple 2D materials with a small unit cell of a few atoms, a $6\times6\times1$ 2D BZ sampling
grid is sufficient to converge the calculated quasiparticle band gap to within 0.02 $\sim$ 0.05 eV, resulting
in a speed-up factor of over two orders of magnitude compared with the conventional uniform sampling approach.
This method, when combined with another method that we developed earlier\cite{Gao2016},
results in a speed-up factor of well over three orders of magnitude for fully converged $GW$ calculations 
for 2D materials. 

To demonstrate the capability and performance of our method, we have carried out fully converged 
$GW$ calculations for monolayer C$_2$N, a recently discovered 2D material with a large unit cell
of 18 atoms, and investigated its quasiparticle band structure in detail. Our calculations not only
provide most converged results but also reveal interesting features of the 
near-edge electronic properties of this interesting 2D material.

With these development, we can carry out fully converged $GW$ calculations for complex and/or large
2D materials with moderate computational resources that are available to most research groups.
We believe that our developments will greatly facilitate future high throughput screening 
of the quasiparticle properties of 2D semiconductors for various applications. 
{Note that our method only works for 2D semiconductors since the dielectric function and the
electron self-energy for 2D metallic systems have different analytical behaviors. 
In addition, capturing the intra-band transitions in metallic systems may still
require a fairly dense $k$-grid.
It would be interesting to find out if current approach can be extended to metallic systems.} 

\begin{center}
{\bf DATA AVAILABILITY}
\end{center}

The datasets generated during and/or analysed during the current study are available from the corresponding author on reasonable request.

\begin{center}
{\bf CODE AVAILABILITY}
\end{center}

The code developed in this work will be made available from the corresponding author after optimization and on  reasonable request.

\begin{center}
{\bf ACKNOWLEDGMENT}
\end{center}

This work is supported by the NSF under Grant Nos. DMR-1506669 and DMREF-1626967.  
P. Z. acknowledges the Southern University of Science and and Technology (SUSTech) 
for hosting his extended visit during spring 2019 when he was on sabbatical.
Work at SUSTech and SHU is supported by National Natural Science Foundation of China 
(Nos. 51632005, 51572167, and 11929401).  W.Z. also acknowledges the support from the 
Guangdong Innovation Research Team Project (No. 2017ZT07C062),
Guangdong Provincial Key-Lab program (No. 2019B030301001), Shenzhen Municipal Key-Lab program (ZDSYS20190902092905285), 
and the Shenzhen Pengcheng-Scholarship Program.
We acknowledge the computational support provided by
the Center for Computational Research at UB,  Beijing Computational Science Research Center,
and the Center for Computational Science and Engineering at SUSTech.

\begin{center}
{\bf AUTHOR CONTRIBUTIONS}
\end{center}

Weiyi Xia was responsible for most of the methodology and code development. Weiwei Gao participated in the early stage of the methodology and code development. Gabriel Lopez-Candales and Yabei Wu carried out some of the calculations and participated in the discussion. Wei Ren and Wenqing Zhang participated in the discussion and provided insightful suggestions. Peihong Zhang was responsible for the original idea and supervised the project. All authors contributed to the manuscript writing.

\begin{center}
{\bf COMPETING INTERESTS}
\end{center}

The authors declare no Competing Financial or Non-Financial Interests.

\bibliographystyle{naturemag}
\bibliography{w2dgw_no_url}

\end{document}